\pdfoutput=1
\documentclass[%
reprint,
superscriptaddress,
nofootinbib,
amsmath,amssymb,
aps,
]{revtex4-1}
\usepackage{graphicx}% Include figure files
\usepackage{bm}% bold math
\usepackage{hyperref}% add hypertext capabilities
\usepackage[utf8]{inputenc}
\usepackage{amsmath,amssymb}
\usepackage[T1]{fontenc} % Use 8-bit encoding that has 256 glyphs
\usepackage{microtype} % Slightly tweak font spacing for aesthetics
\usepackage[english]{babel} % Language hyphenation and typographical rules

\usepackage{booktabs} % Horizontal rules in tables
\usepackage[utf8]{inputenc}
\usepackage{amsmath}
\usepackage{graphicx}
\usepackage{amssymb}
\usepackage{braket,mleftright}
\usepackage{empheq}
\usepackage{subfigure}
\usepackage{stackrel}
\usepackage{soul}
\usepackage{blkarray}
\usepackage{multirow}
\usepackage{amsmath}
\usepackage{physics}
\usepackage{amsfonts}
\usepackage{bm}
\usepackage{bbold} %math& text
\usepackage{color}

\usepackage[utf8]{inputenc} %acentos
\usepackage{amsmath}

\usepackage{enumitem} % Customized lists
\setlist[itemize]{noitemsep} % Make itemize lists more compact

\usepackage{hyperref} % For hyperlinks in the PDF
\usepackage{braket}
\usepackage{verbatim} %multiline comments

\begin{document}

\title{Exploiting Landscape Geometry to Enhance Quantum Optimal Control}

\author{Mart\'{i}n Larocca}
\email{mail to: larocca@df.uba.ar}

\affiliation{Departamento de F\'{i}sica “J. J. Giambiagi” and IFIBA, FCEyN, Universidad de Buenos Aires, 1428 Buenos Aires, Argentina
}
\author{Esteban Calzetta}
\author{Diego Wisniacki}

\affiliation{Departamento de F\'{i}sica “J. J. Giambiagi” and IFIBA, FCEyN, Universidad de Buenos Aires, 1428 Buenos Aires, Argentina
}%

\date{November, 2019}%

\begin{abstract}

The successful application of Quantum Optimal Control (QOC) over the past decades  unlocked the possibility of directing the dynamics of quantum systems. Nevertheless, solutions obtained from QOC algorithms are usually highly irregular, making them unsuitable for direct experimental implementation. In this paper, we propose a method to reshape those unattractive optimal controls. The approach is based on the fact that solutions to QOC problems are not isolated policies but constitute multidimensional submanifolds of control space. This was originally shown for finite-dimensional systems. Here, we analytically prove that this property is still valid in a continuous variable system. The degenerate subspace can be effectively traversed by moving in the null subspace of the hessian of the cost function, allowing for the pursuit of secondary objectives. To demonstrate the usefulness of this procedure, we apply the method to smooth and compress optimal protocols in order to meet laboratory demands.
\end{abstract}

\maketitle

\section{Introduction} \label{Section-Intro}

Recent technological advances related to the manipulation of quantum systems have triggered the development of new quantum-based technology, a.k.a. the second quantum revolution. Proposals for communication, computation and simulation protocols based on quantum mechanical effects \cite{nielsen2002quantum,gis,bib:nori2014,glaser2015training} are nowadays being transformed into reality thanks to the extraordinary capabilities of physical platforms such as ion traps, quantum dots and superconducting qubits \cite{bib:martinis2016,bib:yin2017,bib:lukin2017,Acin_2018}. In addition, particular protocols require exquisite control of external fields. In this context, optimization methods based on Optimal Control Theory were originally put forward in the late 80s and have been strikingly successful at producing high fidelity control protocols \cite{bib:rabitz1988,bib:tannor1993}.

The standard problem in Quantum Optimal Control (QOC) is to find a control field $\omega(t)$ that maximizes a certain objective functional $J[\omega(t)]$, i.e. the probability of reaching a target state or unitary transformation. In general, this is achieved by introducing a parametrization on the control field function, with M control variables, and by performing a gradient descent procedure \cite{brif2010control}. A large number of pioneer simulations and laboratory experiments rapidly indicated that QOC optimization was remarkably easy \cite{rabitz-phystoday}. To understand this situation, the theory of Quantum Control Landscapes (QCL) was developed \cite{bib:rabitz2004}. The QCL is the hypersurface that maps the space of controls into corresponding values of the objective functional. Remarkably, when no constraints are imposed on the control, the QCL is devoid of sub-optimal local maxima. This result was an important first step in explaining the success of QOC procedures but also lead to misleading conclusions, since constraints are inherent to both laboratory and numerical real-world QOC \cite{pechen2012trap,laro2018}. 

More recently, it was shown that, for the state transfer control problem in finite-dimensional quantum systems, the Hessian of the landscape at a solution has an extensive null space and only a finite number of negative eigenvalues \cite{bib:shen2006,rabitz2006topology}. The result is Hamiltonian-independent and there are \textit{at most} $2N-2$ non-zero hessian eigenvalues, where $N$ is Hilbert-space dimension. This is exactly the minimum number of control variables, $M_{min}$, strictly needed for solutions to the control problem to exist. When more control variables are introduced in the control setting, solutions multiply and form connected level-sets of continuously changing control fields that preserve the yield \cite{beltrani, moore2012exploring}. This emergent plurality of solutions produce an almost trap free QCL. Instead, if the number of controls is shrank and approaches $M_{min}$, traps have been shown to dominate the landscape \cite{pechen2012trap,laro2018}. Other types of constraints, like limiting the pulse amplitudes, bandwidth, fluence or protocol duration, impact in a similar fashion. The common approach to solve control problems is then to place hundreds of control variables (many more than needed) and to perform local optimizations in high dimensional landscapes where optimization algorithms work best.

In this paper, we exploit the existence of continuous \textit{level-sets} of solutions to enhance standard protocols produced by typical QOC algorithms. Originally stated for finite dimensional systems \cite{rabitz2006topology}, we show the result is also valid in a continuous variable control problem: the driving of a quantum harmonic oscillator. We focus on the practical consequences it may have on laboratory QOC. Raw QOC solutions are not suited for experimental implementation due to different reasons including high bandwidth or large amplitudes, which makes it hard for the equipment to cope with \cite{smooth1,smooth2,Stefanatos2009pseudospectral,muller2015dressing,laflamme_enhancing,guery2019comparison,mbeng2019optimal}. In this context, we propose to exploit the aforementioned geometric property to reshape optimal protocols in order to meet two common experimental demands: the need to \textit{smooth} or to \textit{compress} optimal control fields. Starting from an unappealing solution produced with standard QOC pulse engineering framework, we launch secondary objective gradient descents constrained to the principal objective optimal level-set. With this method we are able to produce, in a straightforward way, high fidelity solutions that meet laboratory requirements.

This paper is organized as follows. In Section \ref{Section-Model}, we introduce the model, control task and the associated cost functional. Section \ref{Section-Uland} provides analytic proof of the existence of optimal level-sets in the control landscape. We compute the hessian of the cost functional and show that when evaluated at an arbitrary globally-optimal control, there are only two non-zero eigenvalues.  In Section \ref{Section-Mov} we address real-world QCLs, which are naturally constrained to finite dimensional control spaces. We show how to move inside these sets and provide two examples of second objective optimization. Finally, section \ref{Section- conclu} holds the concluding remarks.

\section{Model and Control Problem} \label{Section-Model}

Consider a particle in a one dimensional time-dependent harmonic trap. The evolution is described by the Hamiltonian
\begin{equation}
\hat{H}(t)=\frac{\hat{p}^2}{2m}+\frac{1}{2}m\omega(t)^2\hat{x}^2
\label{ec:sho}
\end{equation}

\noindent where $\hat{x}$ and $\hat{p}$ are the canonical operators, $m$ is the mass of the particle and $\omega(t)$ the time-dependent frequency of the trap. We set $m=\hbar=1$ throughout the paper. Initially, the trap has frequency $\omega(0)=\omega_0$. Suppose we want to open the trap in such a way that, at time $t=T$, its frequency is $\omega(T)=\omega_T<\omega_0$ while the initial and final occupation numbers are the same (as defined from the initial and final Fock basis respectively). These protocols are particularly important in the design of quantum thermal machines \cite{salamon2009maximum,lutz,ste,stef2016minimum,Kosloff_2017}.

Arbitrary protocols produce what is called \textit{quantum friction}. This comes about because by the end of the protocol the Hamiltonian is no longer diagonal in the basis of states with well defined particle number $\hat N(0)=\hat{a}^{\dagger}(0)\hat{a}(0)$. The Hamiltonian can be diagonalized by introducing a new destruction operator through a Bogoliubov transformation \cite{calzetta2008nonequilibrium}

\begin{equation}
\hat{a}(T)=\alpha\hat{a}(0)+\beta\hat{a}^\dagger(0)
\label{bogo}
\end{equation}
\noindent where $\alpha$ and $\beta$ are protocol-dependent complex coefficients satisfying $|\alpha|^2-|\beta|^2=1$. See Appendix \ref{Section bogo} for a detailed description on how to compute these parameters, given a particular protocol. If the initial state is a well defined particle number state (or an incoherent superposition thereof), then the mean particle number at the end, as measured by the operator  $\hat N(T)=\hat{a}^{\dagger}(T)\hat{a}(T)$, will increase 

\begin{equation}
    N(T)=\left\langle \hat{a}^\dagger(T)\hat{a}(T) \right\rangle= N(0)(1+2|\beta|^2)+|\beta|^2
\end{equation}
\noindent Friction limits work extraction, so friction-less evolution is particularly relevant when designing super-adiabatic strokes in the context of quantum heat engines \cite{salamon2009maximum,lutz,ste,stef2016minimum,Kosloff_2017}. This problem has also been addressed within the Shortcuts to Adiabaticity formalism \cite{delcampo2010,del2013shortcuts, guery2019shortcuts, calzetta2018not}. A natural measure for the departure from target friction-less evolution, or simply \textit{infidelity}, is given by
\begin{equation}
    I_{\omega(t)}(T)=|\beta|^ 2
\label{I}
\end{equation}
\noindent so we will use this as our objective. Note that for an optimal protocol with $|\beta|^ 2=0$ the Bogoliubov transformation reduces to

\begin{equation}
    \hat{a}(T)=\hat{a}(0)e^{i\theta}
\label{theta}
\end{equation}
\noindent so the only remaining parameter is the phase factor $\theta$. 

The hyper-surface defined by the functional Eq. (\ref{I}), mapping real-valued functions $\omega(t)$ into real numbers $I_{\omega(t)}$, is referred to as the Quantum Control Landscape (QCL). Minimization of Eq. (\ref{I}) will achieve any one of the infinite $\theta$-parametrized possible friction-less evolutions. In the following section, we will prove solutions to the control problem, that is, global minima of the QCL, form continuous submanifolds of control space. This degeneracy is not related to the freedom in $\theta$ from Eq. (\ref{theta}). For an inspection of $\theta$-specific QCLs, which also present degenerate subspaces of solutions, please refer to Appendix \ref{Section-symp}.

\section{Hessian Analysis} \label{Section-Uland}

In this section we will show that solutions to the unconstrained, infinite-dimensional QCL live on infinite-dimensional submanifolds of control space, with only a finite number of directions heading away from the solution level-set. In order to explore the critical topology of the landscape, let us calculate the change in infidelity produced by a differential variation of the protocol $\omega(t)$, which is given by the inner product of the gradient and the direction of perturbation $\delta \omega(t)$

\begin{equation}
\delta I
\label{dI}=\int_{0}^{T} \nabla I(t) \delta\omega(t) dt \equiv <\nabla I(t),\omega(t)>
\end{equation}{}

\noindent where 

\begin{equation}
    \nabla I(t)=2Re[\nabla\beta(t)\beta^*]
\label{grad_inf}
\end{equation}{}
\noindent In turn, the gradient of $\beta$ is a complex function given by 

\begin{equation}
    \nabla\beta=\beta |f(t)|^2+\alpha^* f^2(t)
\label{grad_beta}
\end{equation}{}
\noindent where $f(t)$ is the solution to the equation of motion, Eq. (\ref{2nd}) in Appendix \ref{Section bogo}. Follow Appendix \ref{Ap db} for the derivation of Eq. (\ref{grad_beta}). Points in the landscape can be classified into critical and non-critical. Non-critical points have non-zero gradient vector $\nabla I$. Movement along the directions that yield $\delta I=0$, those orthogonal to the gradient, preserves fidelity yield. Instead, critical points are special points on the landscape where the movement in \textit{any} direction produces no first order variation, $\delta I=0$. This is because the gradient is identically zero. A second order variation of the objective

\begin{equation}
\label{hess}
\delta^2 I = \int_{0}^{T} dt \int_{0}^{T} dt' \nabla^2 I(t,t') \delta\omega(t)\delta \omega(t')
\end{equation}

\noindent can provide valuable information about the topology of the QCL. The hessian $\nabla^2 I$ is given by

\begin{equation}
\nabla^2 I(t,t') = 2Re[\nabla\beta(t)\nabla\beta^*(t')+\nabla^2\beta(t,t') \beta^*]
\label{He}
\end{equation}

\noindent At a global optimum, where $I_{\omega(t)}\equiv|\beta|^2=0$, the second term vanishes and the expression is reduced to 

\begin{equation}
\nabla^2 I|_{opt}(t,t') = 2\nabla\beta(t)\nabla\beta^*(t')
\label{Heopt}
\end{equation}

\noindent Considering a basis for the space of allowed control functions $\Phi_i(t)$, with  $i=0...\infty$, we can represent the gradient as a vector, where each component is given by the inner product of $\nabla \beta (t)$ with the corresponding basis function

\begin{equation}
\nabla \beta_i =<\nabla \beta (t), \Phi_i(t)>
\label{gr_matrix}
\end{equation}

\noindent Similarly, the hessian can be expressed as an infinite dimensional square matrix, where each entry is given by

\begin{equation}
\nabla^2 I^{opt}_{ i,j} = 2 <\nabla \beta (t), \Phi_i(t)><\nabla \beta^* (t), \Phi_j(t)>
\label{Heopt_matrix}
\end{equation}

\noindent It is clear from Eq. (\ref{Heopt}) that any basis function $\Phi_i$ which is orthogonal to both $\mathrm{Re}[\nabla\beta]$ and $\mathrm{Im}[\nabla\beta]$ will produce a null entry in the $\nabla^2I^{opt}_{i,j}$ matrix. This means that there are at most two directions of decreasing fidelity in the vicinity of any globally optimal point. 

% \noindent If we choose our basis such that

% \begin{equation}
% \label{dI}
%   \begin{aligned}
%   \Phi_0 &= Re[\nabla\beta(t)] 
%  \\
% \Phi_1 & =  Im[\nabla\beta(t)] 
% \\
% \Phi_{i>1} &  \perp Span(\Phi_0 ,\Phi_1, ... , \Phi_{i-1} )
% \end{aligned}
% \end{equation}

% \noindent it can be readily seen that the Hessian matrix, at a solution, is reduced to a 2x2 matrix, being zero any entry where $i>1$ or $j>1$. This means there are at most two directions of decreasing fidelity in the vicinity of any globally optimal point.

\section{Navigation in Solution Sets and Secondary Objective Optimization.—} \label{Section-Mov}

\begin{figure}
	\begin{center}
		\includegraphics[width=.5\textwidth]{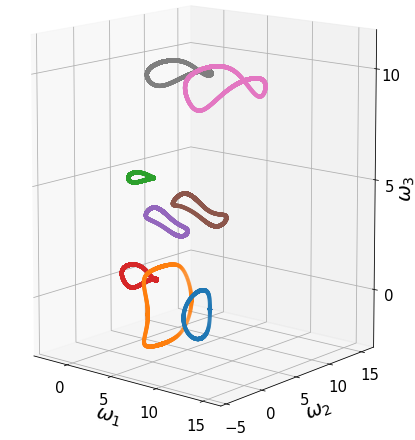}
		\caption{Each point in 3D space represents a possible protocol, where $\omega_i$ the amplitude of the $i_{th}$ pulse in the sequence. Solutions are scattered coloured dots. They form continuous curves, one-dimensional submanifolds of 3D parameter space. An expansion task with $\omega(0)=1$, $\omega(T)=0.25$ and $T=1.8$ was chosen.}
		\label{fig:nts3}
	\end{center}
\end{figure}

In this section, we deal with finite-length piece-wise constant controls. Real-world computer simulations and laboratory experiments are, as mentioned before, intrinsically constrained to finite-length control protocols. Consider a control function $\omega(t)$ represented by a vector of control variables,

\begin{equation}
\omega(t)\rightarrow \left\{\omega_k\right\}\equiv \vec{\omega}
\label{piece}
\end{equation}
\noindent one for each constant-amplitude equal-length time step $dt_k=t_{k+1}-t_k=dt$, where $k=1,2,\ldots,M$. For the control problem considered we have shown,  in the previous section, that the hessian evaluated at an optimal protocol has an extensive null space with only two non-zero eigenvalues. This implies that, if $M>M_{min}=2$ parameters are used, continuous level-sets of solutions arise. To get a better understanding of this situation, in Fig. \ref{fig:nts3}, we draw solution sets for $M=3$. Eight different closed sets of solutions are depicted in the image. We have chosen, for the simulations, an expansion stroke with $\omega(0)=1$, $\omega(T)=0.25$ and $T=1.8$.

An important practical consequence, unfolding from the existence of continuous submanifolds of solutions to the control problem, is that secondary objective gradient descents, constrained to these optimal sets, can be easily put forward. We will tackle two specific demands in present day QOC: (i) we will show how noisy protocols can be leveled-up to produce smooth, experiment-friendly control fields, and (ii) we will demonstrate how long, high-dimensional control sequences can be compressed into equally-optimal low-dimensional solutions.

\begin{figure}
	\begin{center}
		\includegraphics[width=.5\textwidth]{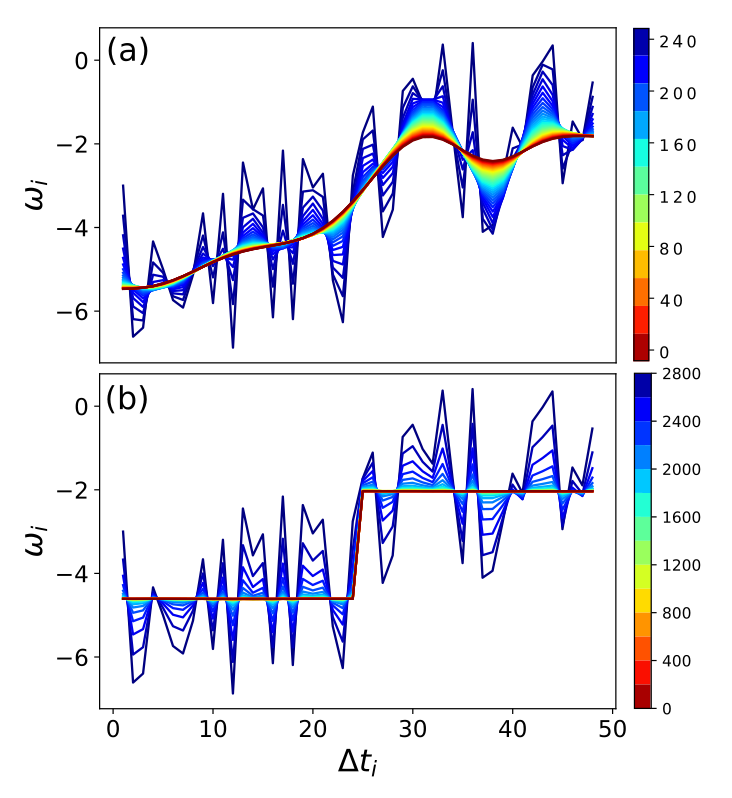}
		\caption{Navigating in solution space to achieve secondary features. Each point in the graph depicts the value of the $i_{th}$ component of a given protocol, corresponding to the time interval $\Delta t_i$. That is, the curves represent distinct protocols, which were coloured relative to their secondary objective cost value. All of them are optimal w.r.t. the main objective in Eq. (\ref{I}). \textbf{(a)} A smoothing descent with $C_1$, and \textbf{(b)} a compression procedure with $C_2$. Both processes start with the same (dark blue) protocol and achieve remarkably different optimal fields (dark red) depending on which secondary cost was targeted.}
		\label{dip}
	\end{center}
\end{figure}

First, let us describe the smoothing procedure. We start from an initial random field and perform a descent following the gradient of $\vert\beta\vert^2$, until a critical point in the landscape is met. If the infidelity is above a certain threshold, $I>I_{th}=10^{-5}$, we choose a new random field and descend again, repeating until a solution is obtained. Please refer to Appendix \ref{Ap num} for details on the numerical methods employed. Once we have a solution, we introduce an auxiliary cost function

\begin{equation}
    C_1=\sum_{i=2}^{M} (\omega_i-\omega_{i-1})^2
\end{equation}{}

\noindent that penalizes jumps between consecutive pulses in the control sequence. Moving in the direction of the gradient of $C_1$ will produce ever-smoother fields. Thus, we initiate a second descent, starting from the original solution and following, this time, the gradient of this auxiliary cost projected into the null subspace of the Hessian, Eq. (\ref{Heopt}). To exemplify, we set $M=48$ and present the achieved trajectory in Fig. \ref{dip} (a). The initial solution, darkest blue curve in the figure, displays a highly irregular profile with $C_1=243$. This control is continuously morphed until the projected gradient vanishes, reaching the dark red curve, with $C_1=0.7$. Although this is the best $C_1$ value for the level-set, it does not necessarily mean it is the best possible smooth solution. Very much like in Fig. \ref{fig:nts3}, where several disconnected solution level-sets exist, it may happen that the smallest values of $C_1$ reside in a different set than the starting one. Since the algorithm is local, it wont be able to reach those values. Nevertheless, although the $M=3$ sets look disconnected, higher-dimensional $M=6$ paths between them were found to exist. With that in mind, coming back to our $M=48$ problem, we take the final (dark red) solution and double the entries of the field to produce a $96$-dimensional one. Now, we are able to continue the descent since the projected gradient is no longer zero in this higher dimensional space. We attain cost values of $C_1=0.03$. This observation suggests local aspects of highly-constrained QCL's may be progressively lost when constraints are released, providing a way of accessing ever-better values of secondary objective.

As second example of this harnessing of the degeneracy of solutions to QOC problems, we demonstrate how protocol compression can be achieved. We contemplate the possibility of compressing an originally M-dimensional field into $L=M/K$ dimensions, with integer $L,K<M$.  The secondary descent is performed with a new cost function that splits the field into $L$ chunks and penalizes the jumps between all of the pulses contained in each chunk

\begin{equation}
    C_2=\sum_{k=1}^{L}\sum_{i>j=1}^{K} (\omega^k_i-\omega^k_j)^2
\end{equation}{}

\noindent where the superscript $k$ indicates which of the $L$ chunks the pulse belongs to and the subscripts $i,j$ span the $K$ pulses inside each chunk. In particular, we will be interested in casting the $M=48$ solution into $L=M_{min}=2$ dimensions. Fig. \ref{dip} (b) presents the compression sequence, starting again with the same (dark blue) 48-dimensional field used to test the smoothing procedure, this time yielding a secondary cost of $C_2=2780$, and achieving a perfect $M=2$ (dark-red) protocol, with $C_2=0$. 

% In highly constrained situations, the sets are disconnected island-like objects. But, when the constraints are released, they tend to merge in one big submanifold. This is important because second objective descents are restricted to one set, and it may be the case that the desired auxiliary features are better met on another set. This local aspect may seem to threaten the performance of the method and the initialization of a large number of random seeds may appear the only way to fully explore the possibilities in the landscape. 

% This is not the case. The \textit{islands} happen to be connected by higher dimensional bridges which allow for the algorithm to access and explore other regions of control space. In higher dimensions, the local characters of the landscape, traps and disconnected sets, seem to vanish and local algorithms achieve globally optimal results.

\section{Summary and Outlook.—} \label{Section- conclu}

In this work, we have exploited an extraordinary property hold by QCLs, namely that optima are not isolated points in the landscape but form continuously varying level-sets that allow for secondary objective optimization. We chose a driven quantum harmonic oscillator to explore the validity of this result, which was originally formulated for discrete systems \cite{rabitz2006topology}, in a continuous variable setting.

Solutions provided by standard QOC techniques are usually not suited for direct experimental use \cite{smooth1,smooth2}, and there is an ongoing effort on designing methods to build protocols that meet laboratory demands \cite{laflamme_enhancing}. We have shown that the exploitation of QCL's geometry can provide a powerful and systematic tool to address those requirements. To illustrate this, we focused on two main experimental challenges. First, we showed how to smooth originally irregular QOC solutions, allowing for the experimental hardware to cope with. In a second example, we demonstrated how long optimal sequences can be compressed into their minimal length partners. Minimal length protocols are important because they minimize the resources needed to control. Any given experimental setup has a natural limit to the maximum number of pulses it can implement. This is due to the fact that there are hardware limitations both to the fastest switching time $\delta T_{min}$, and to the maximum protocol duration $T_{max}$ before decoherence comes into play. Therefore, protocol length is constrained to $M<M_{exp}=T_{max}/\delta T_{min}$. In turn, control experiments are restricted to a maximum system size, $N_{max}=\frac{M_{exp}+2}{2}$. That is, provided the $M_{min}$ protocols can be efficiently generated. Of course, compressed protocols could be searched-for within the standard QOC framework, but previous work has indicated the complex nature of highly constrained landscapes, featuring large trap populations  \cite{pechen2012trap,laro2018}. An approach based on the compression of easy-to-get high dimensional optimal fields, bypassing optimization in trap-signed landscapes, may enable the production of otherwise unattainable solutions. Also, compressed protocols may expose the relevant physical mechanisms employed by the control, in contrast to the obscure nature of high dimensional protocols. 

Quantum Control Landscapes were originally put forward to investigate the complexity of protocol search. Although originally claimed to be trap free, traps have been shown to exist in the most relevant situations, that is, when constraints are imposed on the control. An efficient way to cope with trap dominated landscapes is still one of the main challenges in QOC \cite{bib:sherson2016,bib:sherson2017}, and in optimization theory in general. Aside from their original aim, the study of QCL's can also serve for other purposes. In particular, understanding that solutions live in continuous level-sets allows for the design of methods to achieve secondary features in the controls, as we have shown. A natural extension of the method to more complex scenarios, based on numerical approximations of the Hessian, is treated in \cite{larocca2020navigating}.

\begin{acknowledgements}
M.L. thanks Facundo Sapienza for the fruitful discussions and insights shared. The authors acknowledge financial support
from ANCyPT (Grant no. PICT-2016-1056), CONICET (Grants No PIP 11220170100817CO and PIP 11220150100493CO),
and UBACyT (Grants No 20020130100406BA and 20020170100234BA).

\end{acknowledgements}

\appendix

\section{THE BOGOLIUBOV COEFFICIENT}\label{Section bogo}

In this Section, we explain how to relate control protocols $\omega(t)$ with the infidelity values of Eq. (\ref{I}). To do that, we need an expression for the Bogoliubov coefficient $\beta$. In fact, we don't need to calculate $\alpha$ because it only accounts for the phase factor between the bases, $\theta$. The reader is invited to follow Appendix \ref{Ap db}, where we incorporate such phase factor into the control task.

The general evolution of the system is captured by the equation of motion

\begin{equation}
\label{2nd}
\ddot{f}(t)+\omega^2(t)f(t)=0
\end{equation}

\noindent whose solution, $f(t)$, given initial conditions $ f(0)=\frac{1}{\sqrt{2\omega_0}}, \dot{f}(0)=-i\sqrt{\frac{\omega_0}{2}}$, allows us to relate time-evolved canonical operators in terms of \textit{initial} creation and destruction operators

\begin{equation}
\label{ec:x}
  \begin{aligned}
  x(t)=f(t)a+f^*(t)a^\dagger
 \\
  p(t)=\dot{f}(t)a+\dot{f}^*(t)a^\dagger
  \end{aligned}
\end{equation}

\noindent We will adopt the notation $A\equiv a(T)$, $a\equiv a(0)$. Similarly, imposing final-time conditions to (\ref{ec:sho}), another solution, $g(t)$, relating canonical operators with \textit{final} time second-quantization operators is obtained

\begin{equation}
\label{ec:x2}
  \begin{aligned}
  x(t)=g(t)A+g^*(t)A^\dagger
 \\
  p(t)=\dot{g}(t)A+\dot{g}^*(t)A^\dagger
 \end{aligned}
\end{equation}

\noindent where $g(T)=\frac{1}{\sqrt{2\omega_T}}$ and $\dot{g}(T)=-i\sqrt{\frac{\omega_T}{2}}$. Also, note that $\beta$ measures the degree of non-commutation between initial and final time bases

\begin{equation}
    \beta^*=[a,A]
\end{equation}

To link $[a,A]$ with the solutions to the equation of motion, $f(t)$ and $g(t)$, we can compute the commutator between position and momentum operators written in terms of both bases, (\ref{ec:x}) and (\ref{ec:x2}), with the RHS of (\ref{bogo})

\begin{equation}
\label{sup}
  \begin{aligned}
  f(t)[a,A]+f^*(t)[a^\dagger,A]=-g^*(t)
 \\
  \dot{f}(t)[a,A]+\dot{f}^*(t)[a^\dagger,A]=-\dot{g}^*(t)
  \end{aligned}
\end{equation}

\noindent Finally, multiplying the first equation by $\dot{f}^*$, the second one by $f^*$ and subtracting, we can solve for $[a,A]$. Noting that the Wronskian between $f(t)$ and $f^*(t)$ is $f\dot{f}^*-\dot{f}f^*=i$ (this condition follows directly from the preservation of the canonical commutation relations) we arrive at

\begin{equation}
\beta=-\frac{i}{\sqrt{2 \omega_T}}[\dot{f}(T)+i\omega_T f(T)]
\label{beta}
\end{equation}

\noindent In summary, to calculate the performance of any given protocol: (i) integrate equation (\ref{2nd}) for f(t), (ii) time differentiate $f(t)$ to get $\dot{f}$, (iii) evaluate both at time $T$ and (iv) plug in expression (\ref{beta}). An expression for $\alpha$ follows similarly from Eq. (\ref{sup})

\begin{equation}
    \alpha=\frac{i}{\sqrt{2 \omega_T}}[\dot{f}(T)-i\omega_T f(T)]
\end{equation}{}

\section{THE FAMILY OF $\theta$-PARAMETRIZED LANDSCAPES}\label{Section-symp}

In this Section, we explore a more specific class of control problems in which, besides asking the field to generate a friction-less evolution, we specify the phase, $\theta$, between initial and final bases in Eq. (\ref{theta}). To do so, we briefly introduce some notions on the symplectic formalism. Let $z(t)=(x(t),p(t))^{T}$ denote the quadrature vector and S(t) denote the symplectic matrix associated with a propagator U(t) s.t.

\begin{equation}
    U^\dagger(t) z_{\alpha}(t) U(t) = \sum_{\beta} S_{\alpha\beta}(t) z_{\beta}(t)
\end{equation}

\noindent The time-evolution of the quadrature vector is given by
\begin{equation}
    z(t)=S(t)z(0)
\label{z}
\end{equation}

\noindent  Combining Eqs.(\ref{ec:x}) and (\ref{ec:x2}), we can express S(t) as

\begin{equation}
   S(t)=\sqrt{\frac{\omega_0}{2}}  \left[ {\begin{array}{cc}   f(t)+f^*(t) & \frac{i}{\omega_0}(f(t)-f^*(t)) \\
   \dot{f}(t)+\dot{f}^*(t) & \frac{i}{\omega_0}(\dot{f}(t)-\dot{f}^*(t)) \\  \end{array} } \right]
\end{equation}{}

Now, regarding the objective, it is customary to express the objective in terms of the Frobenious norm between final-time symplectic matrix $S(T)$ and a desired target $W(\theta)$

\begin{equation}
I[\theta]=tr[(S(T)-W(\theta))(S(T)-W(\theta))^T]
\label{lc2}
\end{equation}

To complete the derivation, lets find an expression for the target. Introducing Eq. (\ref{theta}) into a time $t=T$ version of Eq. (\ref{ec:x2}), where $g(T)=\frac{1}{\sqrt{2\omega_T}}$ and $\dot{g}(T)=-i\sqrt{\frac{\omega_T}{2}}$, we get

\begin{equation}
\label{ec:x3}
  \begin{aligned}
  x_{targ}=\frac{1}{\sqrt{2\omega_T}}ae^{i\theta}+\frac{1}{\sqrt{2\omega_T}}a^{\dagger}e^{-i\theta}
 \\
  p_{targ}=-i\sqrt{\frac{\omega_T}{2}}ae^{i\theta}+i\sqrt{\frac{\omega_T}{2}}a^{\dagger}e^{-i\theta}
 \end{aligned}
\end{equation}

\noindent and having in mind

\begin{equation}
    \left( {\begin{array}{cc}
x_{targ} \\
p_{targ} \\
\end{array} } \right)=W(\theta)
\left( {\begin{array}{cc}
x_0 \\
p_0 \\
\end{array} } \right)
\end{equation}{}

\noindent where 

\begin{equation}
\label{ec:x0}
  \begin{aligned}
  x_0=\frac{1}{\sqrt{2\omega_0}}(a+a^\dagger)
 \\
  p_0=-i\sqrt{\frac{\omega_0}{2}}(a-a^\dagger)
  \end{aligned}
\end{equation}

\noindent we arrive at an expression for the family of target matrices

\begin{equation}
    W(\theta)=\sqrt{\frac{\omega_0}{4\omega_T}}e^{i\theta}
\left[ {\begin{array}{cc}
1+e^{-2i\theta} & \frac{i}{\omega_0}(1-e^{-2i\theta}) \\
-i\omega_T(1-e^{-2i\theta}) & \frac{\omega_T}{\omega_0}(1+e^{-2i\theta}) \\
\end{array} } \right]
\end{equation}{}

\noindent Although this is the traditional approach to continuous variable control, we opt in the main text to get rid of this freedom of phase using a $\theta$-independent objective, Eq. (\ref{I}). We explicitly tested the existence of level-sets of solutions in any landscape belonging to the one-dimensional family of landscapes in Eq.(\ref{lc2}). Thus, the degeneracy of solutions in the QCL is not linked to this freedom.

\section{THE GRADIENT OF $\beta$}\label{Ap db}
In order to derive an expression for $\nabla \beta (t)$, let us examine how an infinitesimal variation of the control field perturbs the Bogoliubov coefficient (\ref{beta})
\begin{equation}
\label{db}
  \begin{aligned}
\delta\beta &=i[g(T)\delta\dot{f}(T)-\delta f(T)\dot{g}(T)] \\ &= \int_{0}^{T} [\beta |f(t)|^2+\alpha^* f^2(t)] \delta\omega(t) dt 
  \end{aligned}
\end{equation}

\noindent The functions $\delta f(t)$ and $\delta \dot{f}(t)$ were obtained by solving the inhomogeneous equation

\begin{equation}
    \delta \ddot{f} +\omega^2(t) \delta f = -2f \delta  \omega(t) \omega(t) 
\end{equation}

\noindent with the green function method
\begin{equation}
\delta f = -\int_0^T G_{ret}(t,t')f(t')\delta \omega(t') dt'
\end{equation}

\noindent being $G_{ret}(t,t')=-i[f(t)f^*(t')-f^*(t)f(t')]$. Associating the infinitesimal variation of $\beta$ with an inner product $<\nabla \beta (t), \delta \omega(t)>$, we arrive at the expression for the gradient of $\beta$, Eq. (\ref{grad_inf}).

\section{NUMERICAL METHODS}\label{Ap num}

Let us discuss here the details of the numerical methods applied in the simulations. To determine the evolution of the system, given a particular time-dependent protocol for the trap $\omega(t)$,  Eq. (\ref{ec:sho}) has to be integrated for $f(t)$. In principle, $f(t)$ could be approximated using numerical integration methods. But, since we restrict to piece-wise constant controls, c.f. Eq (\ref{piece}), $f(t)$ and $\dot{f}(t)$ can be computed $exactly$ by concatenating the exact solutions to each constant-amplitude time step

\begin{equation}
\ddot{f_i}(t)+\omega_i f_i(t)=0
    \label{sho disc}
\end{equation}{}

\noindent for $i=1,2,\dots,M$. That is, for $t \in [t_i,t_i+dt]$ we have

\begin{equation}
\label{fi}
\begin{aligned}
f (t) = f_i (t- t_i) \\ \dot f(t) = \dot f_i(t-t_i) 
\end{aligned}
\end{equation}

\noindent where $f_i(t)$ and $\dot{f}_i(t)$ are solutions to the $i_{th}$ Eq. (\ref{sho disc}) with the final value of the solutions to the previous interval as initial conditions

\begin{equation}
\label{fi2}
\begin{aligned}
f_i (t) = f_{i-1} \cos( \omega_i t) + \frac{\dot{f}_{i-1}}{\omega_i} \sin( \omega_i t)
 \\  \dot  f_i (t) =-\omega_i f_{i-1} \sin( \omega_i t) + \dot  f_{i-1}\cos( \omega_i t)  
\end{aligned}
\end{equation}

\noindent Here, $f_0=\frac{1}{\sqrt{2\omega_0}}$ and $\dot{f}_0=-i\sqrt{\frac{\omega_0}{2}}$, and we have adopted the notation $f_{i-1}=f_{i-1}(dt)$ and $\dot{f}_{i-1}=\dot{f}_{i-1}(dt)$. Finally, plugging $f(T)=f_M$ and $\dot{f}(T)=\dot{f}_M$ in Eq. (\ref{beta}) for $\beta$, we compute the infidelity of any given protocol.

To search for solutions, we perform gradient descents in the $M$-dimensional space of controls. Note that the expression for the gradient given in Eq. (\ref{grad_inf}) is not suitable for this purpose, since it would map our piece-wise constant $M$-dimensional protocols onto the infinite-dimensional space of continuous functions. A discrete $\nabla I$, with components

\begin{equation}
\label{gr_disc}
\begin{aligned}
\nabla I_i &= \frac{\partial I}{\partial\omega_i}
 \\ &= 2Re[\nabla\beta_i\beta^*]
\end{aligned}
\end{equation}

\noindent can be obtained from a discrete $\nabla \beta$

\begin{equation}
\label{grbet}
\begin{aligned}
\nabla \beta_i &= \frac{\partial\beta}{\partial\omega_i} \\ &= -\frac{i}{\sqrt{2\omega_T}}[\frac{\partial\dot{f}_M}{\partial\omega_i}+i\omega_T \frac{\partial f_M}{\partial\omega_i}]
\end{aligned}
\end{equation}

\noindent which, in principle, could be computed by building symbolic expressions for
\begin{equation}\label{grbet3}
\begin{aligned}
f_M=f_M(\omega_1, \dots ,\omega_M)  
\\ \dot{f}_M=\dot{f}_M(\omega_1, \dots ,\omega_M)  
\end{aligned}
\end{equation}

\noindent and taking the derivatives w.r.t. each time-step. As a matter of fact, these expressions grow exponentially large with $M$, rendering it impractical for our purposes. We will take an iterative approach instead, just like we did for $f(t)$. Consider the following matrices

\begin{equation}\label{mats}
\begin{aligned}
df = \begin{bmatrix} 
    \frac{\partial f_1}{\partial\omega_1} &  &  \\
    \vdots & \ddots & \\
    \frac{\partial f_M}{\partial\omega_1} &        & \frac{\partial f_M}{\partial\omega_M} 
    \end{bmatrix}\\
d\dot{f} = \begin{bmatrix} 
    \frac{\partial \dot{f}_1}{\partial\omega_1} &  &  \\
    \vdots & \ddots & \\
    \frac{\partial\dot{f}_M}{\partial\omega_1} &        & \frac{\partial\dot{f}_M}{\partial\omega_M} 
    \end{bmatrix}
\end{aligned}
\end{equation}

\medskip
\medskip\noindent where each element, $[df]_{ij}$ corresponds to the derivative of $f_i(t)$ w.r.t. $\omega_j$, and again, it is understood that each function is evaluated at final time $\frac{\partial f_i}{\partial\omega_j}=\frac{\partial f_i}{\partial\omega_j}(dt)$. Of course, these are lower triangular matrices, since $\frac{\partial f_i}{\partial\omega_j}=0$ for $j>i$. Notice that the last row in each matrix is everything we need to compute $\nabla \beta$. To build the matrices, first observe that the diagonal elements can be easily computed by differentiating Eq. (\ref{fi})

\begin{equation}
\begin{split}
     \frac{\partial f_i}{\partial\omega_i}  &= - dt[  f_{i-1} \sin (\omega_i dt) \\
    & -\frac{\dot{f}_{i-1}}{\omega_i} \cos (\omega_i dt) ] -  \frac{\dot{f}_{i-1}}{\omega^2_i}\sin (\omega_i dt) \\
    \frac{\partial\dot{f}_i}{\partial\omega_i}  &= -dt[\omega_i  f_{i-1} \cos (\omega_i dt)+ \dot{f}_{i-1} \sin (\omega_i dt)] \\ & -  f_{i-1} \sin (\omega_i dt)  
\end{split}{}
\end{equation}{}

\noindent These elements can be used, in turn, to compute the sub-diagonal ones

\begin{equation}\label{grbet3}
\begin{aligned}
\frac{\partial f_{i}}{\partial\omega_{i-1}}=\frac{\partial f_{i-1}}{\partial\omega_{i-1}} \cos(\omega_i dt)+\frac{\partial\dot{f}_{i-1}}{\partial\omega_{i-1}}\frac{\sin(\omega_i dt)}{\omega_i}  \\ \frac{\partial\dot{f}_i}{\partial\omega_{i-1}}=-\omega_i \frac{\partial f_{i-1}}{\partial\omega_{i-1}} \sin(\omega_i t)+ \frac{\partial\dot{f}_{i-1}}{\partial\omega_{i-1}}\cos(\omega_i t)
\end{aligned}
\end{equation}

\noindent and with these, those below. Iterating this procedure we are able to build the matrices. A similar approach was taken to construct the exact discrete Hessian, $\nabla^2 \beta_{ij}=\frac{\partial^2 \beta}{\partial\omega_i \partial\omega_j}$, this time using 3-dimensional arrays to represent $d^2f$ and $d^2\dot{f}$.

\bibliography{laro2019.bib}

\end{document}